\documentstyle[12pt]{article}
\catcode`\@=11
\def\marginnote#1{}
\newcount\hour
\newcount\minute
\newtoks\amorpm
\hour=\time\divide\hour by60
\minute=\time{\multiply\hour by60 \global\advance\minute
by-\hour}\edef\standardtime{{\ifnum\hour<12
\global\amorpm={am}%
        \else\global\amorpm={pm}\advance\hour by-12 \fi
        \ifnum\hour=0 \hour=12 \fi
        \number\hour:\ifnum\minute<10
0\fi\number\minute\the\amorpm}}
\edef\militarytime{\number\hour:\ifnum\minute<10
0\fi\number\minute}

\def\draftlabel#1{{\@bsphack\if@filesw {\let\thepage\relax
   \xdef\@gtempa{\write\@auxout{\string
      \newlabel{#1}{{\@currentlabel}{\thepage}}}}}\@gtempa
   \if@nobreak \ifvmode\nobreak\fi\fi\fi\@esphack}
        \gdef\@eqnlabel{#1}}
\def\@eqnlabel{}
\def\@vacuum{}
\def\draftmarginnote#1{\marginpar{\raggedright\scriptsize\tt#1}}
\def\draft{\oddsidemargin -.5truein
        \def\@oddfoot{\sl preliminary draft \hfil
        \rm\thepage\hfil\sl\today\quad\militarytime}
        \let\@evenfoot\@oddfoot \overfullrule 3pt
        \let\label=\draftlabel
        \let\marginnote=\draftmarginnote

\def\@eqnnum{(\theequation)\rlap{\kern\marginparsep\tt\@eqnlabel}%
\global\let\@eqnlabel\@vacuum}  }

\def\numberbysection{\@addtoreset{equation}{section}
        \def\theequation{\thesection.\arabic{equation}}}

\def\underline#1{\relax\ifmmode\@@underline#1\else
 $\@@underline{\hbox{#1}}$\relax\fi}

\catcode`@=12
\relax

\numberbysection
\def\beq{\begin{equation}}
\def\eeq{\end{equation}}
\def\bea{\begin{eqnarray}}
\def\eea{\end{eqnarray}}
\def\p{\partial}
\def\l{\label}
\def\no{\nonumber}

\begin{document}

\begin{titlepage}
\nopagebreak
\begin{flushright}
LPTENS-95/20\\
hep--th/9505047
\\
May 1995
\end{flushright}

\vglue 2.5  true cm
\begin{center}
{\large\bf
HIGHER GRADING GENERALISATIONS  \\
OF THE TODA SYSTEMS\footnote{partially supported by the E.U.
network ``Capital Humain et Mobilit\'e'' contract \#
CHRXCT920069} }
\vglue 1  true cm
{\bf Jean--Loup GERVAIS} and {\bf Mikhail V.  SAVELIEV}\footnote{
On leave of absence from the Institute for High Energy Physics,
142284, Protvino, Moscow region, Russia.
}\\
{\footnotesize Laboratoire de Physique Th\'eorique de
l'\'Ecole Normale Sup\'erieure\footnote{Unit\'e Propre du
Centre National de la Recherche Scientifique,
associ\'ee \`a l'\'Ecole Normale Sup\'erieure et \`a l'Universit\'e
de Paris-Sud.},\\
24 rue Lhomond, 75231 Paris C\'EDEX 05, ~France.}

\medskip
\end{center}

\vfill
\begin{abstract}
\baselineskip .4 true cm
\noindent
In the present paper we obtain some integrable generalisations of the
Toda system generated by  flat connection  forms
  taking values in higher
${\bf Z}$--grading subspaces of a simple Lie algebra,
and construct their
general solutions. One may think of our systems as describing
some  new  fields   of the matter type
 coupled to the standard Toda systems.
This is of special interest in nonabelian Toda theories where the
latter involve  black hole target space metrics.
  We also give a derivation of our conformal system
on the base of the Hamiltonian reduction of the WZNW model;
and discuss a relation between abelian and nonabelian systems generated
by a gauge transformation that maps  the first grading description
to the second. The latter involves grades larger than one.

\end{abstract}
\vfill
\end{titlepage}
\baselineskip .5 true cm
\section{Introduction}
At this time it is hardly necessary to emphasize the importance of
Toda systems. In particular,  much activity is devoted to the solution
of affine Toda theories. Although, the coming discussion may be
applied straightforwardly to this case, we will be explicitly concerned
only  with the non
affine case, i.e. with generalizations of Toda systems associated with
simple  finite dimensional Lie algebras. In this connection
it was shown in our paper \cite{GS92}, that nonabelian Toda theories provide
exactly solvable conformal systems in the presence of a
black hole\footnote{For more recent developments in this direction
see ref.\cite{B94}.}. They
correspond to gauged WZNW models with a nilpotent gauge group,
and are thus
basically different from the ones considered at
 the begining of the nineties following
Witten, see e.g., \cite{W91}, \cite{MSW91}, a review \cite{HS93},
and
references therein. The next, quite natural step, is to extend this
 approach so
to obtain more general black matter configurations
in the presence of some
reasonable matter fields, in attempting
to describe, in particular, black hole formation
from collapsing matter and evaporation. There were a number of papers,
see
e.g., a recent review \cite{Str95} and references therein, where such
configurations have
been discussed. Moreover, as it was shown in some of them,
e.g., \cite{A92}, \cite{BC93}, \cite{RST93}, the corresponding systems
containing the conformal metric factor, the dilaton,
and scalar matter fields,
become solvable (in fact, of the simplest -- Liouville type) for a
special choice of the coupling constants. In the general context of
two--dimensional exactly integrable conformal systems, the arising of
configurations with a more general target space metric and additional
matter fields seems to be very  promising and important, and also very
natural. At the same time, having an experience with a study of  black holes
on the base of nonabelian
Toda theories, we realised that to incorporate matter fields in the game, one
should deal with some relevant generalisation of these theories which possesses
field components of a different nature. In a sense, the nearest, while not
precise and still far--away--standing analogue of such a picture, can be
observed from a consideration of the systems associated with the subspaces
${\cal G}_{-1} \oplus {\cal G}_{-1/2}
\oplus {\cal G}_{0} \oplus {\cal G}_{+1/2} \oplus {\cal G}_{+1}$ of a simple
Lie algebra ${\cal G}$ endowed with a half--integral (${\bf Z}_2$--) gradation,
see e.g., \cite{L83}, \cite{LS92}, \cite{O'R92}, \cite{GO'RRS93}; and from
a component form of the supersymmetric Toda equations associated with
a superalgebra \cite{LSS86}. This is why we decided to turn to a problem of
constructing a generalisation of the Toda system associated with higher
grading subspaces of a ${\bf Z}$--graded simple Lie algebra ${\cal G}$. The
most general form of two--dimensional systems generated by
a flat connection with values in an arbitrary graded Lie algebra, has been
discussed in \cite{LS92} and references therein. In this, the consistency
and integrability of the
equations result from the grading condition imposed on the connection form,
which implies the general solution of the system in a form of holomorphic
factorisation under the condition of finiteness of the growth
of the corresponding Lie algebra.
However, for the systems generated by higher grading subspaces, the given
formulation seems to be too abstract, see equations (\ref{e0.1}) --
(\ref{e-M.1}), and for our aims we need to have  a more precise form of
the equations (desirely, maximally close to the Toda systems (\ref{e0.41}))
and their solutions, which takes into account the properties of a simple
Lie algebra and its
gradation.\footnote{Note that for the case of the area--preserving
diffeomorphisms on $T^2$, this project is recently realised in \cite{S95}
in a form of some higher grading integrable generalisation of the continuous
Toda system. It seems that these equations can be relevant for a description
of matter fields with a background of the ``heavenly'' target space, in
particular, under an appropriate contraction procedure where
 the matter fields
do not react on the Toda fields.}

The standard Toda theories are obtained by requiring that
there exists a gauge, where the underlying flat connection has
non zero components only in appropriate grading subspaces of a
simple Lie algebra endowed with a ${\bf Z}$--gradation,
with grades zero,  one  and minus one.
In the present paper we give a derivation of a system which generalises
the Toda theories (\ref{e0.41}) by some additional functions arising from
the values of the flat connection form in higher grading subspaces.
 These new fields,
as it can be seen from the given expression for the effective action, can
be interpreted as matter fields coupled to the usual Toda system.
  There is a
limit, namely a relevant specialisation the In\"on\"u--Wigner
contraction, where the back reaction of these new fields to the Toda
ones goes to zero, so that they simply propagate in the
fields generated by the standard  Toda solutions. Our general
and rather explicit systems
are  illustrated by examples  of the simple Lie
series $A_r$ endowed with the principal gradation, and the algebra $B_3$
endowed with a nonprincipal gradation. Then,  using an
appropriate modification of the method given in \cite{LS92}, which includes,
in fact, some novel ingredients in comparison with the integration scheme
for the Toda system, we obtain the general solution of the Goursat (boundary
value) problem for our system determined
 by the necessary number of arbitrary
functions. The expression for the solution gives a generalisation of the
tau--function responsable for the Toda system. Here, following a line of
\cite{GO'RRS93}, we give also a derivation of our conformal system on the base
of the Hamiltonian  reduction \cite{O'R92} of the WZNW model.
The conformal properties of the system result applying the stress--energy
tensor method of \cite{BG89}. It is interesting to note that here the
corresponding $W$--elements arise as decomposition coefficients
of the connection form in the so--called $W$--gauge, for a definition
see \cite{GO'RRS93}. After that we discuss a relation
between abelian and nonabelian systems. In terms of the associated
$W$--algebras, our reasonings mean, in particular, that, starting from the
$W$--algebra corresponding to the maximal number of the grading subspaces
for the principal gradation of ${\cal G}$, one can obtain the  $W$--algebras
associated with systems for other gradations, just imposing a relevant
differential polynomial constraints. In terms of the associated $W$--geometries
\cite{GM93}, \cite{GS93}, it means that the target space corresponding to the
system generated by the maximal number of the grading subspaces for the
principal gradation of ${\cal G}$, contains all possible types of the metrics
associated with nonabelian systems, and so all possible types of the black
matter of the type under consideration here, are described by this system.
As an illustration of the form of the $W$--elements for nonabelian systems,
we represent examples of  the $2$nd order for a gradation of the series
$B_r$ when the subalgebra ${\cal G}_0$ is isomorphic to the sum of $r-1$ copies
of ${\it gl}(1)$ and ${\it sl}(2, {\bf C})$ subalgebra; and briefly discuss
with them the problem of the integrability of the corresponding nonabelian
system on a half--line. Finally, by an example of the series $A_r$ endowed
with the principal gradation, we obtain a B\"acklund type transformation
for our system, and give a construction of the corresponding $W$--elements.

\section{Derivation of the Equations}
Let ${\cal M}$ be a two--dimensional manifold with local coordinates
$z_{\pm}$, and ${\cal G}=\mbox{ Lie }G=\oplus_m{\cal G}_m$ be a ${\bf
Z}$--graded simple Lie algebra. Consider a flat connection $A$ in the trivial
fibre bundle ${\cal M}\times G\rightarrow {\cal M}$ with $A_+$- and
$A_-$-components taking values, respectively, in the subspaces $\oplus
_{m=0}^{m_+}{\cal G}_m$ and $\oplus_{m=0}^{m_-}{\cal G}_{-m}$ of the ${\cal
G}$.
Then the zero curvature condition written in a form
\beq
{}[\p_++E^+_0+\sum_{m=1}^{m_+}E^+_m, \p_-+E^-_0+\sum_{m=1}^{m_-}E^-_m]=0,
\l{zc}
\eeq
with ${\cal G}$-algebra valued functions $E^{\pm}_m(z_+,z_-)\in {\cal
G}_{\pm m}$, gives a nonlinear integrable system of partial differential
equations which, for simplicity, we consider for $m_+=m_-=M$ and with a gauge
corresponding to $E^+_0=0$. Namely,
\bea
& & \p_+E^-_0 + \sum_{m=1}^M[E^+_m, E^-_m]=0;\l{e0.1}\\
& & \p_+E^-_m + \sum_{n=1}^{M-m-1}[E^+_n, E^-_{m+n}]+
[E^+_{M-m}, E^-_M]=0;\l{e+m.1}\\
& & \p_+E^-_M=0;\l{e+M.1}\\
& & \p_-E^+_m+[E^-_0, E^+_m] + \sum_{n=1}^{M-m-1}[E^-_n, E^+_{m+n}]+
[E^-_{M-m}, E^+_M]=0;\l{e-m.1}\\
& & \p_-E^+_M+[E^-_0, E^+_M]=0;\l{e-M.1}
\eea
where $1\leq m\leq M-1$. This is just the system written in
\cite{LS92}, and we are going to represent it in a form maximally
close to the Toda system.

Without loosing generality, one can take the function $E^-_0$ in a form
\[
E^-_0=g_0^{-1}\p_-g_0,\]
where $g_0(z_+, z_-)$ is a function on the complex span of the Lie group
$G_0$ corresponding to the subalgebra ${\cal G}_0$. Then, since
\[
\p_-E^+_m+[E^-_0, E^+_m]\equiv  g_0^{-1}\p_-(g_0E^+_mg_0^{-1})g_0,\]
let us rewrite (\ref{e-m.1}) and (\ref{e-M.1}) as
\bea
& & \p_-(g_0E^+_mg_0^{-1})+\sum_{n=1}^{M-m-1}[g_0E^-_ng_0^{-1},
g_0E^+_{m+n}g_0^{-1}]\no\\
& &+[g_0E^-_{M-m}g_0^{-1}, g_0E^+_Mg_0^{-1}]=0;\l{e-m.2}\\
& & \p_-(g_0E^+_Mg_0^{-1})=0.\l{e-M.2}
\eea
Up to inessential transformation, equalities (\ref{e+M.1}) and (\ref{e-M.2})
give
\[
E^-_M=X^-, \quad E^+_M=g_0^{-1}X^+g_0;\]
where $X^{\pm}$ are some constant elements of the subspaces ${\cal G}_{\pm M}$.
With account of these relations, equations (\ref{e0.1}), (\ref{e+m.1}) and
(\ref{e-m.1}) take the form
\bea
& & \p_+(g_0^{-1}\p_-g_0)+[g_0^{-1}X^+g_0, X^-]+
\sum_{m=1}^{M-1}[E^+_m, E^-_m]=0; \l{e0.3}\\
& & \p_+E^-_m + \sum_{n=1}^{M-m-1}[E^+_n, E^-_{m+n}]+
[E^+_{M-m}, X^-]=0;\l{e+m.3}\\
& & \p_-(g_0E^+_mg_0^{-1})+\sum_{n=1}^{M-m-1}[g_0E^-_ng_0^{-1},
g_0E^+_{m+n}g_0^{-1}]\no\\
& & +[g_0E^-_{M-m}g_0^{-1}, X^+]=0.\l{e-m.3}
\eea

With the notations\footnote{By convention, here and in what follows
derivatives act only on the
first term of products, unless parentheses indicate
otherwise.}
\[
E^+_m=\p_+W^+_m, \quad E^-_m=g_0^{-1}\p_-W^-_mg_0,\]
the above written equations\footnote{Of course the letter
$W$ does not mean that
these fields are the conserved $W$ generators.}
 can be represented as follows:
\bea
& & \p_+(g_0^{-1}\p_-g_0)=[X^-, g_0^{-1}X^+g_0]+
\sum_{m=1}^{M-1}[g_0^{-1}\p_-W^-_mg_0, \p_+W^+_m]; \l{e0.4}\\
& & \p_+(g_0^{-1}\p_-W^-_mg_0)=[X^-, \p_+W^+_{M-m}]\no\\
& & +\sum_{n=1}^{M-m-1}[g_0^{-1}\p_-W^-_{m+n}g_0, \p_+W^+_n];\l{e+m.4}\\
& & \p_-(g_0\p_+W^+_mg_0^{-1})=[X^+, \p_-W^-_{M-m}]\no\\
& & +\sum_{n=1}^{M-m-1}[g_0\p_+W^+_{m+n}g_0^{-1}, \p_-W^-_n];\l{e-m.4}
\eea
$1\leq m\leq M-1$.

Looking at these relations, we see that we have two types of equations.
First, (\ref{e0.4}) is similar to the usual Toda equation (except that
$X_\pm$ are fixed elements\footnote{This type of theory was considered
recently for the affine case in \cite{FMG94}.}
in ${\cal G}_{\pm M}$, instead of
${\cal G}_{\pm 1}$), apart from the last term.
 Second, the other equations involve first and higher powers  of
the additional fields, and thus are new. One may visualise our system as
describing a set  of matter fields (the fields $W_m^{\pm}$)
coupled to  Toda fields  (contained in $g_0$). From
this viewpoint, the last term of (\ref{e0.4}) may be regarded
as the back reaction of the additional fields to the Toda
fields. This picture is substanciated by finding
 a procedure that brings
this back reaction to zero for part or all of the
matter fields, so that they simply propagate in the field of a Toda
solution.
This procedure
 is just a special type of the In\"on\"u--Wigner contraction
where  elements of the simple Lie algebra ${\cal G}$ under
consideration are multiplied by  constant parameters, some of which
tend to infinity (or to zero, it depends on a definition) in a consistent
way. It is clear that with such a limit procedure we end up with a
non--semisimple Lie algebra.  In particular, one sees that if the grading
subspaces ${\cal G}_m$ are rewritten  as $\lambda_m{\cal G}_m^{\lambda}$,
and then some $\lambda$'s tend to infinity, then, due to the gradation
property, $[{\cal G}_m, {\cal G}_n]\in {\cal G}_{m+n}$, the procedure is
consistent only with a correct choice of the singular parameters
sequence. Namely, one can, in particular, provide absence of the last $M-1$
terms in (\ref{e0.4}).
Of course, there is a number of other meaningfull
possibilities; some of them are discussed in section 6. The given
conclusion can be equaly reached  from the expression for the effective
action (\ref{eff2.M})  presented below.

In the simplest special case when $M=1$, i.e., the connection takes
values only in the local part ${\cal G}_{-1}\oplus {\cal G}_{0}\oplus
{\cal G}_{+1}$, these equations are reduced to the Toda system \cite{LS92}
\beq
\p_+(g_0^{-1}\p_-g_0)=[X^-, g_0^{-1}X^+g_0]; \l{e0.41}
\eeq
where, if the gradation is consistent with some ${\it sl}(2, {\bf C})$
subalgebra of ${\cal G}$, $X^{\pm}$ can be considered as the nilpotent
elements of this subalgebra.

Next, for $M=2$ we arrive at the system
\bea
& & \p_+(g_0^{-1}\p_-g_0)=[X^-, g_0^{-1}X^+g_0]+
[g_0^{-1}\p_-W^-_1g_0, \p_+W^+_1]; \l{e0.42}\\
& & \p_+(g_0^{-1}\p_-W^-_1g_0)=[X^-, \p_+W^+_1];\l{e+.42}\\
& & \p_-(g_0\p_+W^+_1g_0^{-1})=[X^+, \p_-W^-_1];\l{e-.42}
\eea
the last two equations of which can be reduced to
\beq
g_0^{-1}\p_-W^-_1g_0=[X^-, W^+_1];\quad g_0\p_+W^+_1g_0^{-1}=[X^+, W^-_1].
\l{e+-.42}
\eeq
Then, as it was mentioned in the introduction,  the system
 (\ref{e0.42}),  (\ref{e+-.42})
looks very similar to those based on  half--integral
gradations  of ${\cal G}$, and to
 the supersymmetric Toda system associated with
a superalgebra, however, with a different meaning of the functions
$W^{\pm}_1$ which are there odd functions (with anti--commuting values).

Consider one more explicit example, namely $M=3$,
\bea
& & \p_+(g_0^{-1}\p_-g_0)=[X^-, g_0^{-1}X^+g_0]\no\\
& & +[g_0^{-1}\p_-W^-_1g_0, \p_+W^+_1]
+[g_0^{-1}\p_-W^-_2g_0, \p_+W^+_2]=0; \l{e0.43}\\
& & \p_+(g_0^{-1}\p_-W^-_1g_0)=[X^-, \p_+W^+_2]
+[g_0^{-1}\p_-W^-_2g_0, \p_+W^+_1];\l{e+1.43}\\
& & \p_+(g_0^{-1}\p_-W^-_2g_0)=[X^-, \p_+W^+_1];\l{e+2.43}\\
& & \p_-(g_0\p_+W^+_1g_0^{-1})=[X^+, \p_-W^-_2]
+[g_0\p_+W^+_2g_0^{-1}, \p_-W^-_1];\l{e-1.43}\\
& & \p_-(g_0\p_+W^+_2g_0^{-1})=[X^+, \p_-W^-_1];\l{e-2.43}
\eea
where we can also remove derivatives over $z_+$ and $z_-$ in equations
(\ref{e+2.43}) and (\ref{e-2.43}), respectively,
\beq
g_0^{-1}\p_-W^-_2g_0=[X^-, W^+_1];\quad g_0\p_+W^+_2g_0^{-1}=[X^+, W^-_1].
\l{e+-.43}
\eeq
and, with the help of (\ref{e+-.43}), represent equations (\ref{e+1.43}) and
(\ref{e-1.43}) as
\bea
& & \p_+(g_0^{-1}\p_-W^-_1g_0)=[X^-, \p_+W^+_2]
+[[X^-, W^+_1], \p_+W^+_1];\l{e+1.53}\\
& & \p_-(g_0\p_+W^+_1g_0^{-1})=[X^+, \p_-W^-_2]
+[[X^+, W^-_1], \p_-W^-_1].\l{e-1.53}
\eea
The analogous procedure allows one to bring system (\ref{e0.4}) --
(\ref{e-m.4}) in a form containing multiple commutators of the functions
$W^{\pm}$ (and their derivatives) with the elements $X^{\mp}$.

Let us give also another form of writing this system. Namely, introduce the
functions $V^{\pm}_m=g_0^{\pm 1}\p_{\pm}W^{\pm}_mg_0^{\mp 1}$, with which our
system looks as follows:
\bea
& & \p_+(g_0^{-1}\p_-g_0)=[X^-, g_0^{-1}X^+g_0]+
\sum_{m=1}^{M-1}[V^-_m, g_0^{-1}V^+_mg_0]; \l{e0.5}\\
& & \p_+V^-_m=[X^-, g_0^{-1}V^+_{M-m}g_0]
 +\sum_{n=1}^{M-m-1}[V^-_{m+n}, g_0^{-1}V^+_ng_0];\l{e+m.5}\\
& & \p_-V^+_m=[X^+, g_0V^-_{M-m}g_0^{-1}]
 +\sum_{n=1}^{M-m-1}[V^+_{m+n}, g_0V^-_ng_0^{-1}].\l{e-m.5}
\eea

Note that in the given derivation of the system, we did not specify any
particular ${\bf Z}$--gradation at all. Of course, in the same way as for a
particular case of the Toda system ($M=1$), a choice of a concrete ${\bf
Z}$--gradation allows one to represent the system in terms of equations where
the unknowns are coordinates of the corresponding homogeneous space.  For the
most simple case -- the principal gradation, the equations become essentially
more explicit. Recall that for this gradation the local part ${\cal G}_{-1}
\oplus {\cal G}_0\oplus {\cal G}_{+1}$ is parametrised by the Cartan and the
Chevalley elements $h_j$ and $X_{\pm j}$, $1\leq j\leq r$, corresponding to
the simple roots $\pi_i,\, 1\leq i\leq r$, satisfying the defining relations
\beq
{}[h_i, h_j]=0,\; [h_i, X_{\pm j}]=\pm k_{ji}X_{\pm j}, \; [X_i,
X_{-j}]=\delta_{ij}h_i;\l{dr}
\eeq
where $r$ is the rank of a simple Lie algebra ${\cal G}$ with the Cartan matrix
$k$.

Let us illustrate the system arising for the case of the principal
gradation by an example of the series $A_r$. Parametrise the group element
\[
g_0=\exp (-\sum_{j=1}^rx_jh_j).\]
Then, for $M=1$ when
\[
X^{\pm}=\sum_{j=1}^r\sqrt{c_j^{(1)}}X_{\pm j},\;
c_i^{(1)}=2\sum_{j=1}^r(k^{-1})_{ij},\]
we immediately come, with some evident
re--notation, to the well--known abelian Toda system
\beq
\p_+\p_-x_i=e^{\rho_i}, \quad \rho_i\equiv \sum_{j=1}^rk_{ij}x_j.\l{at0.1}
\eeq
For $M=2$, the elements $X^{\pm}$ are
\[
X^{\pm}=\sum_{j=1}^{r-1}\sqrt{c_j^{(2)}}X_{\pm (j,j+1)},\]
with $X_{\pm (j,j+1)}$ being the root vectors corresponding to the
sum of the simple
roots $\pi_j$ and $\pi_{j+1}$,
and $c_i^{(2)}$ are some constants. Parametrising
the functions
\[
W^{\pm}_1=\sum_{j=1}^rf^{\pm}_j(z_+,z_-)X_{\pm j},\]
we have the equations
\bea
& & \p_+\p_-x_i=
 +c_i^{(2)}e^{\rho_i +\rho_{i+1}} +c_{i-1}^{(2)}e^{\rho_i +\rho_{i-1}}
+e^{-\rho_i}\p_+f^+_i\p_-f^-_i;
\l{at0.2}\\
& & \p_{\pm}f^{\pm}_i=\mp e^{\rho_i}(\sqrt{c_{i-1}^{(2)}}f^{\mp}_{i-1}-
\sqrt{c_i^{(2)}}f^{\mp}_{i+1});\l{at+-.2}
\eea
where one should not forget that $c_0^{(2)}=c_r^{(2)}=f^{\pm}_0=f^{\pm}_{r+1}
\equiv 0$, since all of them are absent in our game. Continuing the
consideration, for the last step, $M=r$, where $X^{\pm}=X_{\pm 12\cdots r}
\equiv X_{\pm\mbox{max}}$ are the maximal and minimal roots
of $A_r$, respectively, we introduce the notations
\beq
W^{\pm}_m=\sum_{\alpha\in\Delta^+_m}f^{\pm m}_{\alpha}(z_+,z_-)X_{\pm\alpha};
\l{set}
\eeq
where $\Delta^+_m$ is the set of the positive roots corresponding to the root
vectors in the subspace ${\cal G}_m$, namely
\bea
& & \Delta^+_1=\{ \pi_1,\cdots , \pi_r\},\;
\Delta^+_2=\{ \pi_1+\pi_2,\cdots , \pi_{r-1}+\pi_r\},\;
\Delta^+_3=\{ \pi_1+\pi_2+\pi_3,\no\\
& & \cdots ,\pi_{r-2}+\pi_{r-1}+\pi_r\},\;
\cdots ,\; \Delta^+_{r-1}=
\{ \pi_1+\cdots +\pi_{r-1}, \pi_2+\cdots +\pi_r\};
\no
\eea
i.e.,
\beq
\Delta^+_m=\{  \sum_{i=1}^m\pi_i,\;  \sum_{i=2}^{m+1}\pi_i,
 \cdots ,  \sum_{i=r-m+1}^r\pi_i\};\quad
1\leq m\leq r.\l{Delta}
\eeq
Moreover, sometimes, to be precise, it is suitable to re--denote
$X_{\alpha}$, $\alpha\in\Delta^+_m$, as
\[
X^m_p\equiv X_{\sum_{i=p}^{m+p-1}\pi_i},\; 1\leq p\leq r-m+1;\]
and also put $f^{\pm m}_{\alpha}\equiv f^{\pm m}_p$.

Then the arising equations are written as
\bea
\p_+\p_-x_i & = & e^{\sum_{j=1}^r\rho_j}-\delta_{i1}\Psi^{r-1}_2
-\delta_{ir}\Psi^{r-1}_1\no\\
& + & \sum_{n=1}^{r-2}\sum_{j=i-n+1}^i\Psi^n_j;
\l{at01.r}
\eea
\bea
& & \p_+[e^{-\sum_{j=i}^{m+i-1}(kx)_j}\p_-f^{-m}_i]\l{at+.r4}\\
& = & \delta_{r-m+1,i}\p_+f^{+(r-m)}_1 - \delta_{1,i}
\p_+f^{+(r-m)}_{m+1} \no\\
& - & \sum_{n=1}^{r-m-i+1}
e^{-\sum_{j=i}^{m+n+i-1}(kx)_j}
 \p_-f^{-(m+n)}_{i}\, \p_+f^{+n}_{i+m}\no\\
& + &\sum_{n=1}^{i-1}
e^{-\sum_{j=i-n}^{m+i-1}(kx)_j}
\p_-f^{-(m+n)}_{i-n}\, \p_+f^{+n}_{i-n},\; 1\leq i\leq r-m+1;\no
\eea
\bea
& & \p_-[e^{-\sum_{j=i}^{m+i-1}(kx)_j}\p_+f^{+m}_i]\l{at-.r4}\\
& = -& \delta_{r-m+1,i}\p_-f^{-(r-m)}_1 + \delta_{1,i}
\p_-f^{-(r-m)}_{m+1} \no\\
& + & \sum_{n=1}^{r-m-i+1}
e^{-\sum_{j=i}^{m+n+i-1}(kx)_j}
 \p_+f^{+(m+n)}_{i}\, \p_-f^{-n}_{i+m}\no\\
& - &\sum_{n=1}^{i-1}
e^{-\sum_{j=i-n}^{m+i-1}(kx)_j}
\p_+f^{+(m+n)}_{i-n}\, \p_-f^{-n}_{i-n},\; 1\leq i\leq r-m+1.\no
\eea
Here
\[
\Psi^m_i\equiv e^{-\sum_{s=i}^{m+i-1}\rho_s}\> \> \p_+f^{+m}_i\p_-f^{-m}_i;\]
 in (\ref{at+.r4}) and  (\ref{at-.r4}) the terms on the
3rd line corresponding to $i=1,\, n=r-m$, and the terms on the
4th line corresponding to $i=r-m+1,\, n=r-m$ are evidently absent.
Moreover, of course,  in system (\ref{at01.r}) appear only those functions
$\Psi^n$ which enter decomposition (\ref{set}), in other words
\bea
\p_+\p_-x_i & = & e^{\sum_{j=1}^r\rho_j}+\Psi^1_i
+(\Psi^2_{i-1}+\Psi^2_{i})\no\\
 & + & (\Psi^3_{i-2} +\Psi^3_{i-1}+\Psi^3_{i})\no\\
& + & \cdots \l{at0.r}\\
& + & [(1-\delta_{ir})\Psi^{r-1}_1
+(1-\delta_{i1})\Psi^{r-1}_2].\no
\eea

So, equations (\ref{at01.r}), (\ref{at+.r4}),  and (\ref{at-.r4}) represent
a system generated by the flat connection form with values in all
higher grading subspaces of the Lie algebra $A_r$ endowed with the
principal gradation; its general solution follows from the formulas given
below.

Note that in the case under consideration, the subalgebra ${\cal G}_0$ contains
exactly $m-1$ elements commuting with the subspace ${\cal G}_m$, namely
\bea
h_i,\; r-m+2\leq i\leq m-1;\; & & h_i+h_{m+i},\; 1\leq i\leq r-m;\no\\
& & h_{r-m+1}-\sum_{i=m}^rh_i;\l{ger}
\eea
and  equations (\ref{at01.r}),  (\ref{at+.r4}) and (\ref{at-.r4}) reflect
this property.

\section{Effective Action}
It is known that the effective action for the Toda system (\ref{e0.41})
can be written as
\beq
{\cal I}^{(1)}_{\mbox{eff}}={\cal I}_{\mbox{wz}}(g_0)-\lambda\int_{\cal M}
\mbox{ tr }(X^-g_0^{-1}X^+g_0),\l{eff1}
\eeq
where ${\cal I}_{\mbox{wz}}(g_0)$ is the Wess--Zumino type action, see e.g.
\cite{O'R92}. For system (\ref{e0.42}) -- (\ref{e-.42}), i.e.
for the case with $M=2$, the effective action is written in terms
of the functions $g_0$ and $W^{\pm}_1$, namely
\bea
& & {\cal I}^{(2)}_{\mbox{eff}}={\cal I}_{\mbox{wz}}(g_0)-\lambda\int_{\cal
M}\mbox{ tr }[(X^-g_0^{-1}X^+g_0)\l{eff2.1}\\
& & -(\p_+W^+_1g_0^{-1}\p_-W^-_1g_0)
-([X^-, W^+_1] \p_+W^+_1)-([X^+, W^-_1] \p_-W^-_1)];\no
\eea
or in terms of the functions $g_0$ and $V^{\pm}_1$, however, with an
additional integration.
This form (\ref{eff2.1}) of the action ${\cal I}^{(2)}_{\mbox{eff}}$ coincides,
up to some inessential factors, with those obtained in \cite{O'R92} for a
half--integral gradation of ${\cal G}$. One can see already from this first
example, how it generalises the corresponding expression, e.g. in \cite{Str95},
including the matter fields.

In the general case under consideration here, an effective action corresponding
to system (\ref{e0.3}) -- (\ref{e-m.3}) is written as
\bea
 {\cal I}^{(M)}_{\mbox{eff}}&=&{\cal I}_{\mbox{wz}}(g_0)-\lambda\int_{\cal
M}\mbox{ tr }[(X^-g_0^{-1}X^+g_0)\l{eff2.M}\\
&-&\sum_{m=1}^{M-1}\{ (\p_+W^+_mg_0^{-1}\p_-W^-_mg_0)
+([X^-, W^+_m] \p_+W^+_{M-m})\no\\
&+&([X^+, W^-_m] \p_-W^-_{M-m})\}
 + *** ;\no
\eea
where the symbol $***$ here means the terms of type $([\cdots [X^{\mp},
W^{\pm}_{m_1}], \-W^{\pm}_{m_2}]\-\cdots ]\- \p_{\pm}W^{\pm}_{M-\sum m_i})$.

\section{General Solution}
Let us construct the general solution to system (\ref{e0.4}) --
(\ref{e-m.4}) using the method given in \cite{LS92}. At first,
we repeat briefly the general formulation of the last
reference. This introduces   the necessary
notations in order
to proceed with our problem by specialising that method.

The flat connection $A$ in question is represented in the gradient form,
\beq
A_{\pm}=g^{-1}\p_{\pm}g,\l{cm}
\eeq
with $g\in G$; and we take for $A_{\pm}$ the modified Gauss decomposition
of $g$,
\beq
g=M_-N_+g_{0-} \mbox{ and } g=M_+N_-g_{0+},\l{gauss}
\eeq
respectively. The grading conditions realised in (\ref{zc}), provide the
holomorphic property of $M_{\pm}$, namely that  the functions
$M_{\pm}(z_{\pm})\in G_{\pm}$ satisfy the initial value problem
\beq
\p_{\pm}M_{\pm}(z_{\pm})=M_{\pm}(z_{\pm})L_{\pm}(z_{\pm}),\l{ivp}
\eeq
where
\beq
L_{\pm}(z_{\pm})=\sum_{m=1}^M{\cal E}^{\pm}_m(\Phi^{\pm}); \quad
{\cal E}^{\pm}_m(\Phi^{\pm})=\sum_{\alpha\in\Delta^+_m}\Phi^{\pm m}_{\alpha}
(z_{\pm})X_{\pm\alpha},\l{arb}
\eeq
with arbitrary functions $\Phi^{\pm m}_{\alpha}(z_{\pm})$ determining the
general solution to our system; the meaning of $\Delta^+_m$ and $X_{\pm\alpha}$
is the same as in (\ref{set}). The functions $M_{\pm}(z_{\pm})$
can be represented by the corresponding multiplicative integrals (like the
halved $S$--matrix in the quantum field theory, see e.g. \cite{BSh59}), or
by a noncommutative version of the known exponential formula,
see  \cite{Str87}, \cite{LS92}.

So, with account of (\ref{cm}) and (\ref{gauss}), we have
\beq
A_{\pm}=g_{0\mp}^{-1}(N_{\pm}^{-1}\p_{\pm}N_{\pm})g_{0\mp}+
g_{0\mp}^{-1}\p_{\pm}g_{0\mp}.
\l{A+-}
\eeq
Differentiating the identity
\beq
M_+^{-1}M_-=N_-\hat{g}_0N_+^{-1}, \l{id}
\eeq
with $\hat{g}_0\equiv g_{0+}g_{0-}^{-1}$,
over $z_{\pm}$ and using equations (\ref{ivp}), one can  get convinced
that the elements $N_{\pm}^{-1}\p_{\pm}N_{\pm}$ take values in the subspaces
$\oplus_{m=1}^{M}{\cal G}_{\pm m}$. Rewrite now decomposition
(\ref{A+-}) as
\beq
A_{\pm}=g_{0\pm}^{-1}\tilde{L}_{\pm}g_{0\pm}+
g_{0\mp}^{-1}\p_{\pm}g_{0\mp},\l{AG}
\eeq
where
\[
\tilde{L}_{\pm} \equiv g_{0}^{\pm 1}(N_{\pm}^{-1}\p_{\pm}N_{\pm})g_{0}^{\mp 1}
=
\sum_{m=1}^M{\cal E}^{\pm}_m(F^{\pm})\]
with some unknowns $F^{\pm m}_{\alpha}(z_+, z_-)$. Note that here the functions
$\tilde{L}_{\pm}$ are equal to $L_{\pm}$ only when $M=1$.
However, thanks to identity (\ref{id}), the elements $N_{\pm}$ and $\hat{g}_0$
are determined by the elements $M_{\pm}(z_{\pm})$, and hence the functions
$F^{\pm}(z_+,z_-)$
entering $\tilde{L}_{\pm}$ can be expressed in terms of the functions
$\Phi^{\pm m}_{\alpha}(z_{\pm})$ for an arbitrary $M$; moreover,
it is clear that $F^{\pm M}=\Phi^{\pm M}(z_{\pm})$.

Now, let us accomodate this general consideration to system (\ref{e0.4}) --
(\ref{e-m.4}), or better to (\ref{e0.5}) -- (\ref{e-m.5}), for which the
corresponding connection components have the form
\beq
A_+=g_0^{-1}(\sum_{m=1}^{M-1}V^+_m+X^+)g_0,\quad
A_-=g_0^{-1}\p_-g_0+\sum_{m=1}^{M-1}V^-_m+X^-.\l{AS}
\eeq
Equating decompositions (\ref{AG}) and (\ref{AS}), one has
\bea
& & g_0^{-1}(\sum_{m=1}^{M-1}V^+_m+X^+)g_0=g_{0+}^{-1}\tilde{L}_+g_{0+}+
g_{0-}^{-1}\p_+g_{0-};\l{eq+}\\
& & g_0^{-1}\p_-g_0=g_{0+}^{-1}\p_-g_{0+},\;
\sum_{m=1}^{M-1}V^-_m+X^-=g_{0-}^{-1}\tilde{L}_-g_{0-};\l{eq-}
\eea
thereof
\bea
& & g_0=(y_0^+)^{-1}(z_+)g_{0+},\quad  g_{0-}=y_0^-(z_-);\l{rel1}\\
& &  {\cal E}^-_M(\Phi^-)=g_{0-}X^-g_{0-}^{-1},\quad
g_{0+}^{-1}{\cal E}^+_M(\Phi^+)g_{0+}=
g_{0}^{-1}X^+g_{0}; \l{rel2}\\
& & V^-_m=g_{0-}^{-1}{\cal E}^-_m(F^-)g_{0-},\quad
g_0^{-1}V^+_mg_0=g_{0+}^{-1}{\cal E}^+_m(F^+)g_{0+}.\l{rel3}
\eea
Here $y^{\pm}_0\equiv y^{\pm}_0(z_{\pm})\in G_0$ are arbitrary
functions of their arguments, and are expressed in terms of $\Phi^{\pm}$.
Then it is clear that $g_0=(y_0^+(z_+))^{-1}\hat{g}_0y_0^-(z_-)$.
Rewriting identity (\ref{id}) in the form
\[
(y^+_0)^{-1}M_+^{-1}M_-y^-_0=[(y^+_0)^{-1}N_-y^+_0]g_0
[(y^-_0)^{-1}N_+^{-1}y^-_0],\]
one has
\beq
<h \vert (y^+_0)^{-1}M_+^{-1}M_-y^-_0\vert h' >=<h\vert g_0\vert h' >, \l{hv}
\eeq
where the brackets are taken between the basis vectors $\vert h >$
and the dual, $< h'\vert$, to  $\vert h' >$, annihilated by the subspaces
${\cal G}_+$ and ${\cal G}_-$, respectively. This matrix element realises a
higher grading generalisation of the standard tau--function for the Toda
system (\ref{e0.41}) depending on the necessary number
of arbitrary functions
$\Phi^{\pm m}_{\alpha}(z_{\pm})$ which determine the general solution to
system (\ref{e0.4}) -- (\ref{e-m.4}) or  (\ref{e0.5}) -- (\ref{e-m.5}). It
can be rewritten as a series over the nested integrals of the products of
these functions in the same way as it was done for the case of the
Toda system \cite{LS92} in terms of the basis $\vert h>$ with the help
of the commutation
relations (\ref{dr}). So, formulas (\ref{hv}), (\ref{rel1}) --- (\ref{rel3})
define the general solution to our system.

Note that expression (\ref{hv}) formally is the same as those for the Toda
system (\ref{e0.41}), see \cite{LS92}; however, the functions $L_{\pm}$
entering the initial value problem (\ref{ivp}) for $M_{\pm}$ contain, in
general, contributions coming from higher grading subspaces ${\cal G}_m,\,
1\leq m\leq M$.
The basis vectors $\vert h >$ are the highest vectors of the fundamental
representations of ${\cal G}$ only for the case when the subalgebra ${\cal
G}_0$
is abelian, i.e., for the principal gradation of ${\cal G}$; while for other
${\bf Z}$--gradations the subspace of the vectors annihilated by the action
of ${\cal G}_+$ is not already one--dimensional. At the same time,
the connection form for a nonabelian Toda system (\ref{e0.41}) can be obtained
by an appropriate gauge transformation from a connection form generated by
a relevant number of higher grading subspaces of ${\cal G}$ endowed with
the principal gradation,  for which the group element $g_0$ is determined only
by
the diagonal matrix elements in (\ref{hv}) corresponding to the highest vectors
of
the fundamental representations of ${\cal G}$.

The functions $F^{\pm}_m, \,   1\leq m\leq m_{\pm}-1$, entering solution
(\ref{rel1}) -- (\ref{rel3}), and in turn $\tilde{L}_{\pm}$, are determined
in terms of the matrix elements of the known element $M_+^{-1}M_-$ taken
between some, not necessarily highest vectors $\vert h>$ of the representation
space. Consider, for example, the matrix element with a highest bra-vector
$<h'\vert$ and a ket-vector $\vert h >^{(1)}$ which is annihilated by the
action of the subspaces ${\cal G}_m,\, m>1$. Then,
differentiating the equality
\[
<h'\vert \hat{g}_0^{-1}M_+^{-1}M_-\vert h>^{(1)}= <h'\vert
(\hat{g}_0^{-1}N_-\hat{g}_0)N_+^{-1}\vert h>^{(1)}=
<h'\vert N_+^{-1}\vert h>^{(1)},\]
over $z_+$, one has a sequence of equalities
\bea
\p_+<h'\vert \hat{g}_0^{-1}M_+^{-1}M_-\vert h>^{(1)} & = &
-<h'\vert  g_0^{-1}\tilde{L}_+g_0
N_+^{-1}\vert h>^{(1)}=\no\\
-<h'\vert  g_0^{-1}{\cal E}_1(F^+_1)g_0N_+^{-1}\vert h>^{(1)} & =
& -<h'\vert  g_0^{-1}{\cal E}_1(F^+_1)g_0\vert h>^{(1)},\no
\eea
which determines the functions $F^+_1$. Now, knowing these functions,
to find $F^+_2$ we consider the matrix element with a ket-vector $\vert
h>^{(2)}$ which is annihilated by the action of the subspaces ${\cal G}_m,\,
m>2$; etc. The similar procedure allows to determine the functions $F^-_m$.

\section{Connection with WZNW model}
In this section, we  show how the connection between the usual (grading one)
Toda theory and WZNW model extends to our case. It seems to be important
by itself, and also hopefully clarifies the derivation of the general
solutions given in the above section. For background references
on this aspect of the standard Toda theory, we refer to \cite{O'R92}, while the
present discussion closely  follows the line of \cite{GO'RRS93}.

So far, the general solution was discussed with a special form (\ref{AS}),
i.e.,
\beq
A_+=g_0^{-1} V^+ g_0, \; A_-=V^-+g_0^{-1}\partial_- g_0;\;
V^{\pm}\equiv X^{\pm}+\sum_{m=1}^{M-1}V^{\pm}_m.
\l{Toda}
\eeq
As emphasized for instance in  \cite{GO'RRS93}, it is useful to discuss the
properties of the basic zero curvature condition
\beq
\partial_+A_--\partial_-A_++[A_+,  A_-]=0
\l{zeroc}
\eeq
in various relevant gauges related by transformations of the type
\beq
A_\pm\to A_\pm^{h_0}\equiv h_0^{-1} A_\pm h_0 +
h_0^{-1} \partial_\pm h_0,\l{gtr}
\eeq
where $h_0\in G_0$. From this viewpoint, the discussion was so far
essentially carried out in the so--called Toda gauge where $A_\pm$
take the  specific form displayed in (\ref{Toda}). On the contrary,
(\ref{A+-}) is valid in any gauge, and it is easy to see that $M_\pm$
and $N_\pm$ are gauge invariant functions.
In the Toda gauge, (\ref{A+-}) reduces to (\ref{Toda}) if we choose
$g_{0\pm}$ as in (\ref{rel1}). There is another
Toda gauge,  where the components $A_{\pm}$ are exchanged so
that the corresponding
gauge potential $\tilde A_\pm=A_\pm^{g^{-1}_0}$ takes the form
\beq
\tilde A_-=g_0 V^- g_0^{-1}, \quad \tilde A_+=V^++g_0\partial_+
g_0^{-1}.
\l{Todat}
\eeq
The connection with the WZNW model
is seen by changing gauge to two different choices denoted
$A_\pm^{(W+)}$ and $A_\pm^{(W-)}$, such that $A_+^{(W+)}=0$
and $A_-^{(W-)}=0$, respectively. Writing
\beq
A_\pm^{(W+)}=\omega_+ A_\pm \omega_+^{-1}+\omega_+
\partial_\pm \omega_+^{-1},\quad
A_\pm^{(W-)}=\omega_-^{-1} \tilde A_\pm \omega_-
+\omega_-^{-1}\partial_\pm \omega_-,
\l{AWdef}
\eeq
the condition just recalled implies
\beq
\omega_+^{-1}\partial_+ \omega_+=g_0^{-1}V^+g_0, \qquad
 \omega_-\partial_- \omega_-^{-1}=g_0V^-g_0^{-1}.
\l{ompmdef}
\eeq
The other components give the WZNW currents according to
\bea
A_-^{(W+)}&=&-J_-=\omega_+V^-\omega_+^{-1}+\omega_+
g_0^{-1}\partial_-(g_0\omega_+^{-1}),\no\\
A_+^{(W-)}&=&-J_+=\omega_-^{-1}V^-\omega_-
+\omega^{-1}_-g_0\partial_+(g_0^{-1}\omega_-).
\l{Jdef}
\eea
It follows from the zero curvature condition (\ref{zeroc}) that
$\p_+ J_-=0$, and
$\p_-J_+=0$. Moreover, using (\ref{AWdef}), one verifies
that
\beq
J_-=\p_- \omega \omega^{-1}, \quad
J_+=-\omega^{-1}\p_+\omega; \quad
\l{WZNWeq}
\eeq
where
\beq
\omega=\omega_+g^{-1}_0\omega_-,
\l{omdef}
\eeq
so that $\omega$ is a solution of the WZNW equation. According to (\ref{AS}),
$V^\pm\in \oplus_{m=1}^M{\cal G}_m$, with $V^\pm_{\pm M}=X^\pm$
being  fixed elements in
${\cal G}_{\pm M}$. Making use of relations (\ref{ompmdef}), one easily sees
that, as a consequence, $J_-\in \oplus_{m\geq -M}{\cal G}_m$, $J_+\in
\oplus_{m\leq M}{\cal G}_m$, and
\beq
J_-\vert_{-M}=-X^-, \qquad  J_+\vert_M=-X^+.
\l{confred}
\eeq
Those are the conditions that specify the conformal reduction from the general
WZNW
model, and which generalise the conditions derived in  \cite{O'R92} and
\cite{GO'RRS93}
for the usual (grading one) case. Next, repeating the
calculation of \cite{GO'RRS93}, one
obtains the explicit form of the WZNW solution,
\beq
\omega_+=q_+(z_-)(y_0^-)^{-1} N_+y_0^-, \quad
\omega_-=(y_0^+)^{-1}N_-^{-1}y_0^+q_-(z_+);
\l{omgauss}
\eeq
$$
\omega=g_R^{-1}(z_-)g_L^{-1}(z_+),
$$
\beq
g_R=M_-y_0^-q_+^{-1} ,\quad g_L=q^{-1}_-(y_0^+)^{-1}M^{-1}_+.
\l{sol}
\eeq
Finally,  (\ref{hv}) is recovered by taking matrix elements of $\omega^{-1}$
between highest
weight states, and remembering definition (\ref{omdef}). This gives
\beq
<h| \omega^{-1}|h'>=<h|g_0|h'>=
<h| (y_0^+)^{-1}M_+^{-1} M_-y_0^-|h'>.
\l{zerocomp}
\eeq
The arbitrary functions of one variable which appear in the general solution,
are
contained in $L_\pm$ defined by (\ref{arb}). So, we have
to relate
$y_0^{\pm}$
with them. Repeating the reasonings given below (\ref{id}), we have
$$
L_+=N_-\hat g_0N_+^{-1}\p_+N_+\hat g_0^{-1} N_-^{-1}+
N_-\hat g_0\p_+(\hat g_0^{-1} N_-^{-1}).
$$
Since $L_+\in \cal G_+$, only the first term gives nonzero contribution.
Moreover, it is easily seen that, for the maximum grading component
$$
\left. L_+\right |_M= \hat g_0\left. (N_+^{-1}\p_+N_+)\right|_M\hat g_0^{-1}.
$$
Combining (\ref{A+-}) with (\ref{Toda}), this gives
$$
\left. L_+\right |_M=  y_0^+V^+(y_0^+)^{-1}.
$$
Recalling that the maximum grading component of $V^+$ is fixed, and
performing similar calculations for $L_-$, we arrive at the conditions
\beq
\left. L_\pm\right |_{\pm M}= y_0^{\pm}X^\pm (y_0^{\pm })^{-1}.
\l{Lcond}
\eeq
 For the usual (grading one) case, there is only one grading component,
and $y_{0 \pm}$ contain all
arbitrary functions. As it was already mentioned in the previous section,
this is no more true here, of course. Thus, following what
we discussed in that section, we have to take matrix elements of
$\omega^{-1}$ between states
which are not highest weight vectors. As a result, the factors $q_\pm$
do not disappear, and we need to
express them in terms of the arbitrary functions $\Phi_\alpha^{\pm m}$.
This is done as follows.
Substituting expressions (\ref{sol}) in the equation for $J_-$, one sees that
$$
J_-=\p_-g_R^{-1} g_R=\p_-q_+q_+^{-1}+q_+
\p_-(y_0^-)^{-1}y_0^-q_+^{-1}
-q_+(y_0^-)^{-1} L_- y_0^-q_+^{-1},
$$
where we made use of (\ref{ivp}). It follows from (\ref{arb}) and
(\ref{Lcond}) that  we can write
\beq
L_\pm(z_\pm)=\sum_{m=1}^M    y_0^{\pm}
\tilde \epsilon^\pm_m(z_\pm) (y_0^{\pm})^{-1},
\quad  \widetilde \epsilon^\pm_M=X_\pm,
\l{arbt}
\eeq
so that
\beq
J_-=\p_-q_+q_+^{-1}+q_+\p_-(y_0^-)^{-1}y_0^-q_+^{-1}
-\sum_{m=1}^Mq_+\tilde \epsilon^\pm_m(z_\pm)q_+^{-1}.
\l{Jexp}
\eeq
The first two terms of this relation belong to ${\cal G}_+\oplus  {\cal G}_0$.
One may verify that
$q_+$ can be chosen such that  the current takes the form
$$
J_-=-X_-+\sum_{m\geq 0} J_m,\qquad J_m\in {\cal G}_m.
$$
Indeed, writing
$$
q_+=\exp(\sum_{m>0}\alpha^+_m),\qquad \alpha^+_m\in {\cal G}_m,
$$
one gets the equations
\bea
J_-\vert_{-M}&=&-X_-,\no\\
J_-\vert_{-M+1}&=&-\tilde \epsilon^-_{M-1}-[\alpha^+_1, X_-]=0,\no\\
J_-\vert_{-M+2}&=&-\tilde \epsilon^-_{M-2}-[\alpha^+_2, X_-]-
{1\over 2}[\alpha^+_1, [\alpha^+_1, X_-]] =0, \l{qdef}
\eea
and so on, which may be solved reccursively.
Of course, what we are doing
is just fixing gauge
in the corresponding Drinfeld--Sokolov equation \cite{DS}.

\section{An Example of Nonabelian $B_3$ Toda System
Coupled to Matter Fields}

Recall that the simplest nonabelian Toda system which describes
a nontrivial target space metric, is related to the algebra $B_2$ endowed with
a nonprincipal gradation, see \cite{LS92}, \cite{S90}, \cite{GS92}. However,
the systems which incorporate a nontrivial target space metric and nontrivial
matter fields, are realised for the algebras of higher rank, begining from
$B_3$. Just this case is considered in the present section, for which we give
an explicit form of the equations of motion for the
Toda type and matter type fields; write down the corresponding action; and
discuss, how the In\"on\"u--Wigner contraction brings the back reaction of
the matter fields to the Toda fields to zero.
Consider the Lie algebra $B_3$ endowed with
a nonprincipal gradation, namely such that
\bea
&& {\cal G}_0=\{ h_1, h_2, h_3; X_{\pm 3}\} ;\quad {\cal G}_{\pm 1}=\{ X_{\pm
1},
X_{\pm 2}, X_{\pm 23}, X_{\pm 233}\} ;\l{grad}\\
&& {\cal G}_{\pm 2}=\{ X_{\pm 12},
X_{\pm 123}, X_{\pm 1233}\} ;\quad {\cal G}_{\pm 3}=\{ X_{\pm 12233}\} .\no
\eea
Then system (\ref{e0.5}) -- (\ref{e-m.5})
is rewritten in an explicit way in terms of the group parameters of $g_0(z_+,
z_-)$
and the functions $f^{\pm}_a(z_+, z_-)$ entering the decomposition
$V^{\pm}=\sum_{\alpha\in\Delta_1}f^{\pm}_{\alpha}X_{\pm\alpha}$; here and in
what
follows in this section, we omit the lower indices ``1'' at matter fields, and
refering on system  (\ref{e0.5}) -- (\ref{e-m.5}) suppose that there $M=2$.
Furthermore, seemingly the simplest, though rather nontrivial possibility
arises when
one takes as $X^{\pm}$ the root vectors $X_{\pm 1233}$; a hint for such a
choice comes
from the fact that just this element of ${\cal G}_{\pm 2}$ commutes with the
maximal number
of the elements of ${\cal G}_0$, namely with $h_1+h_2$ and $2h_2+h_3$.

Note that in what follows, to provide a possibility to make a reduction to
the corresponding standard Toda system, we supply the elements $X^{\pm}$
with a constant factor $\xi$ which tends to zero in this case. Moreover,
sometimes it is more suitable to use an analogous, while not completely
equivalent form  (\ref{e0.42}), (\ref{e+-.42})
of equations (\ref{e0.5}) -- (\ref{e-m.5}),
where the functions $v^{\pm}_{\alpha}(z_+, z_-)$
enter the decomposition
$W^{\pm}=\sum_{\alpha\in\Delta}v^{\pm}_{\alpha}X_{\pm\alpha}$
with $V^{\pm}=g_0^{\pm 1}\p_{\pm}W^{\pm}g_0^{\mp 1}$.

Let us begin with a suitable parametrisation of the function $g_0$.
It happens that the parametrisation
\[
g_0=e^{a^+X_{+3}}\cdot e^{a^-X_{-3}}\cdot e^{\sum_{i=1}^3a_ih_i}\]
used in \cite{LS92} and then in \cite{S90}, \cite{GS92}, is not the best one;
much more
relevant for nonabelian systems in question is the following:
\beq
g_0=e^{\alpha h_3}\cdot e^{\beta (X_{+3}+X_{-3})}\cdot e^{\gamma h_3}\cdot
e^{\sum_{i=1}^2a_ih_i}.\l{par}
\eeq
It is clear that the parameters in both of them are connected by the formulas
\[
e^{-a_3}=e^{-(\alpha +\gamma )}\mbox{ ch}\beta ,\quad
a^+=e^{2\alpha}\mbox{ th}\beta , \quad a^-=\frac{1}{2}e^{-2\alpha}
\mbox{ sh}2\beta .\]

Note also that the Wess--Zumino term ${\cal I}_{\mbox{wz}}(g_0)$ in
parametrisation (\ref{par}) is given by the expression
\bea
&&{\cal I}_{\mbox{wz}}(g_0)=\frac{1}{2}\int_{\cal M}\{ 4\p_+\alpha\p_-\alpha
+\p_+(a_2-2\gamma )\p_-(a_2-2\gamma )\l{wzpar}\\
{}&&+\p_+a_1\p_-a_1 + \p_+a_2\p_-a_2+4\p_+
\beta\p_-\beta -4\p_+(a_2-2\gamma )\p_-\alpha\mbox{ch}2\beta\} .\no
\eea

With this parametrisation, system  (\ref{e0.5})-- (\ref{e-m.5}) takes the form
\bea
&&\p_+\p_-a_1=-4\xi e^{-a_1+a_2-2(\alpha +\gamma )}\mbox{ch}^2\beta -
\Phi_{1,1},\l{g0h1}\\
&&\p_+\p_-a_2= -4\xi e^{-a_1+a_2-2(\alpha +\gamma )}\mbox{ch}^2\beta
-\Phi_{2,2} -2\Phi_{23,23} -4\Phi_{233,233},\no\\
&&\p_+(\p_-\gamma +\mbox{ch}2\beta\p_-\alpha )=-4\xi e^{-a_1+a_2-2(\alpha +
\gamma )}\mbox{ch}^2\beta -\Phi_{23,23} -4\Phi_{233,233},\no\\
&&\p_+[e^{a_2-2\gamma}( \p_-\beta +\mbox{sh}2\beta\p_-\alpha )]=
\Phi_{23,2}-2\Phi_{233,23},\no\\
&&\p_+[e^{-a_2+2\gamma}(\p_-\beta -\mbox{sh}2\beta\p_-\alpha )]=
 2\xi e^{-a_1-2\alpha}\mbox{ sh }2\beta
+\Phi_{2,23}-2\Phi_{23,233},\no
\eea
\beq
\p_{\pm}f^{\mp}_1=\mp 4\xi F^{\pm}_{233},\quad \p_{\pm}f^{\mp}_{233}=
\pm \xi F^{\pm}_1,\no
\eeq
\beq
\p_{\pm}f^{\mp}_2=\p_{\pm}f^{\mp}_{23}=0.\l{g1.2.23}
\eeq
Here, for the sake of brevity, we denote $\Phi_{\alpha ,\beta}\equiv
F^+_{\alpha}
f^-_{\beta}$; $F^+_{\alpha}$ are the coefficients in the
decomposition $g_0^{-1}V^+g_0 =\sum_{\alpha\in\Delta}F^+_{\alpha}X_{+\alpha}$;
the similar definitions is taken for the functions $F^-_{\alpha}$ in terms of
$f^-_{\alpha}$. It follows from equations (\ref{g1.2.23})
that $f^{\mp}_2$ and $f^{\mp}_{23}$ do not depend on $z_{\pm}$, respectively,
and,
in fact, these functions can be considered as constants without loosing a
generality.
Note that the equations describing the corresponding standard Toda
system associated with such a nonprincipal gradation of $B_3$, arise
when $\xi$ tends to zero, $f^{\pm}_2=f^{\pm}_{233}=0$, and
$f^{\pm}_1$, $f^{\pm}_{23}$ are some nonzero fixed constants.

To emphasise better the structure of these equations, let us make
a simplification, namely put $f^{\pm}_2=f^{\pm}_{23}=0$. Then system
(\ref{g0h1}) -- (\ref{g1.2.23}) is reduced to the following one:
\bea
&&\p_+\p_-a_1= -4\xi e^{-a_1+a_2-2(\alpha +\gamma )}\mbox{ch}^2\beta -
e^{-2a_1+a_2}f^+_1f^-_1,\l{g0a1}\\
&&\p_+\p_-a_2=-4e^{-2(\alpha +\gamma )}\mbox{ch}^2\beta [\xi e^{-a_1+a_2}+
e^{a_1}f^+_{233}f^-_{233}],\l{g0a2}\\
&&\p_+\p_-\beta =-\frac{1}{2}\mbox{th}\beta\p_+\p_-a_2+
4\frac{\mbox{ch}\beta}{\mbox{sh}^3\beta}
\p_+\omega\p_-\omega ,\l{g0beta}\\
&&\p_+(\mbox{cth}^2\beta\p_-\omega )+\p_-(\mbox{cth}^2\beta\p_+\omega )
=\frac{1}{2}\p_+\p_-a_2,\l{g0omega}\\
&& \p_{\pm}f^{\mp}_1=\mp 4\xi e^{a_1-2(\alpha +\gamma )}\mbox{ch}^2\beta
f^{\pm}_{233},\l{g1233s}\\
&&\p_{\pm}f^{\mp}_{233}=\pm\xi e^{-2a_1+a_2}f^{\pm}_1.\l{g1s}
\eea
Here we have used the new function $\omega$ defined by the relations
\[
\p_-\alpha = \frac{\p_-\omega}{\mbox{sh}^2\beta},\qquad
\p_+(\gamma -\frac{a_2}{2}) = \frac{\p_+\omega}{\mbox{sh}^2\beta};\]
which are noncontradictive due to equations (\ref{g0h1}).
In the limit $\xi =0$ and $f^{\pm}_{\alpha}=0$, these equations take the form
\bea
&&\p_+\p_-a_1=\p_+\p_-a_2= 0,\l{g0a0}\\
&&\p_+\p_-\beta = 4\frac{\mbox{ch}\beta}{\mbox{sh}^3\beta}
\p_+\omega\p_-\omega ,\l{g0beta0}\\
&&\p_+(\mbox{cth}^2\beta\p_-\omega )+
\p_-(\mbox{cth}^2\beta\p_+\omega )=0;\l{g0omega0};
\eea
and correspond to the following Lagrange function:
\[
{\cal L}=\frac{1}{2}\p_+\beta \p_-\beta -2\mbox{ cth}^2\beta
\p_+\omega \p_-\omega .\]
Another simplification of system (\ref{g0a1}) -- (\ref{g1s}), similar
to the last one, and effectively still  without matter fields, can be obtained
equating
$\xi e^{-a_1+a_2}+e^{a_1}f^+_{233}f^-_{233}=0$. Then (\ref{g0a2})
gives $\p_+\p_-a_2=0$; and, up to an appropriate transformation, taking $a_2=0$
and, for simplicity, $\xi =1$, one can choose $f^+_{233}=f^-_{233}=ie^{-a_1}$.
Hence, from (\ref{g1s}) one obtains $f^{\pm}_1=\mp i\p_{\pm}e^{a_1}$, and
from (\ref{g1233s}) it follows that $a_1$ satisfies the equation
\beq
\p_+\p_-e^{a_1}=-4e^{-2(\alpha +\gamma )}\mbox{ch}^2\beta ;\l{j1}
\eeq
which coincides with (\ref{g0a1}) under our assumptions. The rest two equations
(\ref{g0beta}) and (\ref{g0omega}) are rewritten in the standard for the
corresponding nonabelian Toda equations form (\ref{g0beta0}) and
(\ref{g0omega0}).

As it was mentioned in section 2, to provide zero back reaction of the matter
fields to the Toda fields for the systems
associated with higher grading subspaces of a simple Lie algebra ${\cal G}$,
one
can apply the In\"on\"u--Wigner contraction of ${\cal G}$. For the
standard Toda
systems such a procedure was realised in \cite{LS92} and references therein;
here we use it for our
system (\ref{e0.5}) -- (\ref{e-m.5}), or, the same, for (\ref{e0.42}),
 (\ref{e+-.42}).
Multiply the elements of the subspaces ${\cal G}_{\pm 1}$ and ${\cal G}_{\pm
3}$
of the algebra $B_3$ by some parameter $\kappa^{-1}$, keeping the elements of
the rest
subspaces ${\cal G}_0$ and ${\cal G}_{\pm 2}$ as they are, i.e., come to an
isomorphic
(for finite values of $\kappa$)
algebra ${\cal G}^{(\kappa )}=\kappa^{-1}{\cal G}_{\pm 3}\oplus{\cal G}_{\pm
2}\oplus
\kappa^{-1}{\cal G}_{\pm 1}\oplus{\cal G}_0$; and then tend $\kappa$ to
infinity in the
commutation relations for the elements of ${\cal G}^{(\kappa )}$.\footnote{Of
course,
this is only one of all posssible contractions of the algebra $B_3$.} As a
result we arrive
at the algebra ${\cal G}^{(\infty )}$ which is not already simple; it is the
semi--direct sum
of the semisimple Lie algebra
$B_2(=\{ X_{\pm 12}, X_{\pm 123}, X_{\pm 1233}, X_{\pm 3}, h_1+h_2, h_3\}
)\oplus
{\it gl}(1)(=\{ 2h_2+h_3\} )$ and the $10$ dimensional commutative multiplet of
the elements
$R_a^{\pm},\, 1\leq a\leq 5$, obtained from the elements $X_{\pm 1}, X_{\pm 2},
X_{\pm 23},
X_{\pm 233}, X_{\pm 12233}$. With this procedure, system (\ref{e0.5}) --
(\ref{e-m.5})
takes the form
\bea
& & \p_+(g_0^{-1}\p_-g_0)=[X^-, g_0^{-1}X^+g_0],\l{g0c}\\
& & \p_{\pm}V^{\mp}=[X^{\mp}, g_0^{\mp 1}V^{\pm}g_0^{\pm 1}];\l{g1c}
\eea
where equation (\ref{g0c}) formally looks as the standard nonabelian Toda
system associated with the Lie algebra $B_2\oplus {\it gl}(1)$ endowed with the
chosen gradation (of course, the meaning of the elements $X^{\pm}$ is different
than there); while the second one describes an abelian multiplet of the fields
$r_i^{\pm},\, 1\leq i\leq 4$, entering the decomposition
$V^{\pm}=\sum_ir^{\pm}_i
R^{\pm}_i$. In a component form these equations read as
\bea
& & \p_+\p_-a_1= -4e^{-a_1+a_2-2(\alpha +\gamma )}\mbox{ch}^2\beta ,\l{g0h1c}\\
& & \p_+\p_-a_2= -4e^{-a_1+a_2-2(\alpha +\gamma )}\mbox{ch}^2\beta ,\l{g0h2c}\\
& & \p_+(\p_-\gamma +\mbox{ch}2\beta\p_-\alpha )=-4e^{-a_1+a_2-2(\alpha +
\gamma )}\mbox{ch}^2\beta ,\l{g0h3c}\\
& & \p_+[e^{a_2-2\gamma}(\p_-\beta +\mbox{sh}2\beta\p_-\alpha )]=0,\l{g0X+3c}\\
& & \p_+[e^{-a_2+2\gamma}(\p_-\beta -\mbox{sh}2\beta\p_-\alpha )]=
 2e^{-a_1-2\alpha}\mbox{ sh }2\beta  ;\l{g0X-3c}\\
&&\p_{\pm}r^{\mp}_1 =  \mp 4e^{a_1-2(\alpha +\gamma )}[-\frac{1}{2}e^{4\alpha}
\mbox{ sh}^2\beta r^+_2 - \frac{1}{2}e^{2\alpha}\mbox{ sh}2\beta r^{\pm}_3\no\\
&&+\mbox{ ch}^2\beta r^+_4],\qquad
\p_{\pm}r^{\mp}_4  =  \pm e^{-2a_1+a_2}r^{\pm}_1,   \l{g14r}\\
&&\p_{\pm}r^{\mp}_2  = \p_{\pm}r^{\mp}_3= 0.\l{g23r}
\eea
Here, by the evident reason, we omit the factor $\xi$; if it is zero, then
the system is trivialised. Introducing the new functions $2\nu\equiv a_1-a_2,\,
2\sigma\equiv a_1+a_2$, one sees that $\nu$ satisfies the Laplace equation
$\p_+\p_-\nu = 0$, whose general solution is
$\nu (z_+, z_-)=\nu^+ (z_+)+\nu^- (z_-)$ with $\nu^{\pm}(z_{\pm})$ being
arbitrary functions of their arguments. Hence, shifting the functions $\gamma$
and $\sigma$ by $-\nu$, we have the following form of our system:
\bea
& & \p_+\p_-\sigma = -4e^{-2(\alpha +\gamma )}\mbox{ch}^2\beta ,\l{g0af}\\
& & \p_+[\p_-(\gamma -\sigma )+\mbox{ch}2\beta\p_-\alpha ]=0,\l{g0gammaf}\\
& & \p_+[e^{\sigma -2\gamma}(\p_-\beta +\mbox{sh}2\beta\p_-\alpha
)]=0,\l{g0abf}\\
& & \p_+[e^{-\sigma +2\gamma}(\p_-\beta -\mbox{sh}2\beta\p_-\alpha )]=
 2e^{-\sigma -2\alpha}\mbox{ sh }2\beta  ;\l{g0baf}\\
&&\p_{\pm}r^{\mp}_1  =  \mp 4e^{\sigma -2(\alpha +\gamma )}[-\frac{1}{2}
e^{4\alpha}\mbox{ sh}^2\beta r^{\pm}_2
- \frac{1}{2}e^{2\alpha}\mbox{ sh}2\beta r^{\pm}_3\no\\
&&+\mbox{ ch}^2\beta r^{\pm}_4],\qquad
\p_{\pm}r^{\mp}_4  = \pm e^{-\sigma}r^{\pm}_1,\l{g14rcf}\\
&&\p_{\pm}r^{\mp}_2  = \p_{\pm}r^{\mp}_3= 0.\l{g23rcf}
\eea
An effective action corresponding to subsystem (\ref{g0af}) -- (\ref{g0baf})
describing a (nontrivial) target space metric and the dilaton  field, is given
by the expression
\bea
{\cal I}_{\mbox{eff}}^{(2)}&=&\frac{1}{2}\int_{\cal M}\{ 4\p_+\alpha\p_-\alpha
+\p_+(\sigma -2\gamma )\p_-(\sigma -2\gamma )+\p_+\sigma \p_-\sigma
\l{effparc}\\
&+& 4\p_+\beta \p_-\beta -4\p_+(\sigma -2\gamma )
\p_-\alpha\mbox{ch}2\beta +8e^{-2(\alpha +\gamma )}\mbox{ch}^2\beta \} .\no
\eea

With a new function $\omega$ defined by the differential relations
\beq
\p_+\omega =-\mbox{ sh}^2\beta\p_+(\gamma -\frac{\sigma}{2}),
\qquad \p_-\omega =-\mbox{ sh}^2\beta\p_-\alpha ,\l{nonloc}
\eeq
which are not contradictive because system (\ref{g0af}) -- (\ref{g0baf})
provides the compatibility condition, $\p_+\p_-\omega =\p_-\p_+\omega$;
one can easily get convinced that equations (\ref{g0gammaf}) -- (\ref{g0baf})
are rewritten as
\bea
&&\p_+\p_-\beta = \mbox{ sh}2\beta e^{-2(\alpha +\gamma )}+
4\frac{ \mbox{ ch}\beta}{\mbox{ sh}^3\beta}\p_+\omega\,\p_-\omega ,\l{eqbeta}\\
&&\p_+(\mbox{ cth}^2\beta\p_-\omega +\frac{1}{4}\p_-\sigma )+
\p_-(\mbox{ cth}^2\beta\p_+\omega +\frac{1}{4}\p_+\sigma )=0.\l{eqomega}
\eea

It seems natural to consider a subclass of the solutions to equations
(\ref{g14rcf})
-- (\ref{g23rcf}) corresponding to a choice $r^{\pm}_2=r^{\pm}_3=0$ when this
system for the matter fields takes the form
\beq
\p_{\pm}r^{\mp}_1 =  \mp 4e^{\sigma -2(\alpha +\gamma )}\mbox{ ch}^2
\beta r^{\pm}_4,\qquad \p_{\pm}r^{\mp}_4  =  \pm e^{-\sigma}r^{\pm}_1.
\l{redmf}
\eeq
Then, using (\ref{g0af}), the first set of equations (\ref{redmf}) is rewritten
as
\[
\p_{\pm}r^{\mp}_1 =  \pm e^{\sigma}\p_+\p_-\sigma \, r^{\pm}_4;\]
thereof, with account of the second set of equations (\ref{redmf}), we obtain
\beq
\p_{\pm}(e^{-\sigma}r^{\mp}_1)=\pm\p_{\mp}(r^{\pm}_4\, \p_{\pm}\sigma ).
\l{nic1}
\eeq
After that, again with the help of the second set of equations (\ref{redmf}),
we represent functions in the parenthesis in the l.h.s. of (\ref{nic1})
as $e^{-\sigma}r^{\mp}_1=\mp\p_{\mp}r^{\pm}_4$, so that (\ref{nic1})
is read as
\[
\p_{\mp}(\p_{\pm}r^{\pm}_4+r^{\pm}_4\p_{\pm}\sigma )=0.\]
Hence, for the matter fields $r^{\pm}_4$ and $r^{\pm}_1$ we have the
following general solutions in terms of the function $\sigma$ and two
arbitrary functions $\sigma^{\pm}_0$ satisfying the Laplace equation:
\beq
r^{\pm}_4=e^{-\sigma -\sigma^{\pm}_0},\qquad r^{\pm}_1=
e^{\sigma}\p_{\pm}e^{-\sigma -\sigma^{\pm}_0}.\l{solr}
\eeq
The general solution to the Toda type system (\ref{g0af}) -- (\ref{g0baf})
corresponding to the effective action (\ref{effparc}), or the system of form
(\ref{g0af}), (\ref{eqbeta}) and (\ref{eqomega}), is obtained directly
from the general solution for equations (\ref{e0.5}) -- (\ref{e-m.5}),
with the same contraction procedure as above. It is determined by four
(or, respectively, three) arbitrary functions depending on $z_+$ and the same
number of functions of $z_-$; and looks similar to those for the standard
nonabelian Toda system \cite{LS92} associated with the algebra $B_3$
endowed with gradation (\ref{grad}).

\section{Relation Between Abelian and Nonabelian Systems}

Of course, a given Lie algebra posseses different gradations.
So, typically we may consider
\beq
{\cal G}=\oplus_{m\in {\it Z}} {\cal G}_m^1\equiv
{\cal G}_{-}^1\oplus {\cal G}_0^1\oplus {\cal G}_{+}^1=
\oplus_{m\in {\it Z}} {\cal G}_m^2\equiv
{\cal G}_{-}^2 \oplus{\cal G}_0^2\oplus {\cal G}_{+}^2.
\label{difgr}
\eeq
In general, a gauge potential satisfying the  grading
condition  with respect to the first gradation, i.e.
$A_\pm \in {\cal G}_0^1\oplus {\cal G}_{\pm}^1$, will not satisfy
it  with respect to the second, i.e.
$A_\pm \notin {\cal G}_0^2\oplus {\cal G}_{\pm}^2$. However it may be
gauge equivalent to a vector potential which does.
This is the case if the first gradation is of the nonabelian type
-- that is if ${\cal G}_0^1\equiv {\cal G}_0^{\mbox{na}}$ is
non abelian,
while the second is the
principal  gradation where ${\cal G}_0^2\equiv{\cal G}_0^{\mbox{a}}$
is generated by the
Cartan elements.
Thus, in principle, one can relate  this or that nonabelian system
 to  an abelian system
 whose connection form
contains the relevant number of the grading subspaces, which will be
larger than one, but not necessarily maximal.
In this, for the principal gradation of ${\cal G}$,
$\sum_{m=1}^{m_{\mbox{max}}}
\mbox{ dim }{\cal G}_m=\sum_i l_i$ where $l_i$
are the exponents of ${\cal G}$.
Here the corresponding gauge transformation which, of course, changes the
grading spectrum of the initial connection form, is generated by some
nonabelian subalgebra of ${\cal G}_0^{\mbox{na}}$. In terms
of the associated
$W$--algebras, it means that, starting from the $W$--algebra corresponding
to an appropriate   number of the grading subspaces
in (\ref{zc}) for the principal
gradation of ${\cal G}$, one can obtain the  $W$--algebras associated with
the systems like (\ref{e0.41}) for other gradations, just imposing a relevant
differential polynomial constraints. In terms of the associated
$W$--geometries,
it means that the target space corresponding to the system generated by
the maximal number of the grading subspaces for the principal gradation of
${\cal G}$ contains all possible types of the metrics associated with
nonabelian
systems  (\ref{e0.41}) and, in particular,
all possible types of the black matter described
by this system.

Let us give some examples related to the simple Lie algebras $B_2$, $B_3$ and
$B_4$.

{\bf $B_2$}: Here the set $\Delta^+_m$ of the positive roots corresponding to
the
root vectors in the subspace ${\cal G}_m$ for the principal gradation, is
\[
\Delta^+_1=\{ \pi_1,\pi_2\},\;
\Delta^+_2=\{ \pi_1+\pi_2\},\;
\Delta^+_3=\{ \pi_1+2\pi_2\};\]
and to enlarge the (abelian) subalgebra ${\cal G}_0^{\mbox{a}}=\{ h_1, h_2\}$
to a nonabelian one, equiped by the elements ${\cal G}_0^{\mbox{na}}=\{ h_1,
h_2, X_{\pm 1}\}$, it is enough to make a gauge transformation, generated by
the
elements $X_{\pm 1}$, of the connection form (\ref{zc}) taking values in the
subspaces ${\cal G}_{\pm m}$ with $m=1,2$. At the same time, the
analogous procedure related to ${\cal G}_0^{\mbox{na}}=\{ h_1, h_2,
X_{\pm 2}\}$,
requires to use the connection with values in the subspaces with $m=1,2,3$.

Note that here, for the principal gradation, the subalgebra ${\cal G}_0$
contains $1$ element
($h_1$) commuting with ${\cal G}_3$, and $1$ element ($h_2$) commuting with
${\cal G}_2$.

{\bf $B_3$}: Here the set $\Delta^+_m$ for the principal gradation, is
\[
\Delta^+_1=\{ \pi_1,\pi_2,\pi_3\},\;
\Delta^+_2=\{ \pi_1+\pi_2, \pi_2+\pi_3 \},\;
\Delta^+_3=\{ \pi_1+\pi_2+\pi_3, \pi_2+2\pi_3\};\]
\[
\Delta^+_4=\{ \pi_1+\pi_2+2\pi_3\}; \;
\Delta^+_5=\{ \pi_1+2\pi_2+2\pi_3\};\]
and to enlarge the (abelian) subalgebra ${\cal G}_0^{\mbox{a}}=\{ h_1, h_2,
h_3\}$ to a nonabelian one, equiped by the elements ${\cal G}_0^{\mbox{na}}=
\{ h_1,h_2, h_3, X_{\pm 1}\}$, it is enough to make a gauge transformation,
generated
by the elements $X_{\pm 1}$, of the connection form (\ref{zc}) taking values in
the subspaces ${\cal G}_{\pm m}$ with $m=1,2,3$. The analogous
procedure related to ${\cal G}_0^{\mbox{na}}=\{ h_1, h_2, h_3, X_{\pm 2}\}$,
requires
to use the connection with values in the subspaces with $m=1,2,3,4,5$; and
those
for ${\cal G}_0^{\mbox{na}}=\{ h_1, h_2, h_3, X_{\pm 3}\}$, requires
the connection with values in the subspaces with $m=1,2,3,4$.

Here, for the principal gradation, the subalgebra ${\cal G}_0$ contains $2$
elements
($h_1, \, h_3$) commuting with ${\cal G}_5$; $2$ elements ($h_1+h_2, \,
2h_2+h_3$)
commuting with ${\cal G}_4$; $1$ element ($h_2$) commuting with ${\cal G}_3$;
and
$1$ element ($h_1+h_2+h_3$) commuting with ${\cal G}_2$.

{\bf $B_4$}: Here the set $\Delta^+_m$ for the principal gradation, is
\[
\Delta^+_1=\{ \pi_1,\pi_2,\pi_3,\pi_4\},\;
\Delta^+_2=\{ \pi_1+\pi_2, \pi_2+\pi_3, \pi_3+\pi_4 \};\]
\[
\Delta^+_3=\{ \pi_1+\pi_2+\pi_3, \pi_2+\pi_3+\pi_4, \pi_3+2\pi_4\},\]
\[
\Delta^+_4=\{ \pi_1+\pi_2+\pi_3+\pi_4, \pi_2+\pi_3+2\pi_4\}; \]
\[
\Delta^+_5=\{ \pi_1+\pi_2+\pi_3+2\pi_4, \pi_2+2\pi_3+2\pi_4\};\]
\[
\Delta^+_6=\{ \pi_1+\pi_2+2\pi_3+2\pi_4\};
\Delta^+_7=\{ \pi_1+2\pi_2+2\pi_3+2\pi_4\}.\]

Here the subalgebra ${\cal G}_0$ contains $3$ elements
($h_1, \, h_3,\, h_4$) commuting with ${\cal G}_7$; $3$ elements ($h_1-h_3, \,
h_2+h_3,\, h_4$) commuting with ${\cal G}_6$; $2$ elements ($h_2,\, h_1+h_3$)
commuting with ${\cal G}_5$; $2$ elements ($h_2+h_3,\, 2h_3+h_4$)
commuting with ${\cal G}_4$; $1$ element ($2h_1+2h_2+h_4$) commuting with
${\cal G}_3$; and $1$ element ($h_2+h_3+h_4$) commuting with ${\cal G}_2$.

To clarify better the picture, give as an illustration a component form of
equations (\ref{e0.41}) which are obtained by the gauge transformation
generated by the elements $X_{\pm r}$ from those for the principal
gradation of the algebra $B_r$ with account of its higher grading subspaces.
The arising system looks as follows \cite{LS92}:
\bea
& & \p_+\p_-x_i=e^{(kx)_i}-(\delta_{i,r-1}-\delta_{i,r})
\frac{\mbox{ sh }\frac{x_{r-1}-x_r}{2}}{\mbox{ ch }^3\frac{x_{r-1}-x_r}{2}}
\p_+x_{r+1}\p_-x_{r+1};\l{bh}\\
& & \p_+\p_-x_{r+1}=
 -\frac{\p_-(x_{r-1}-x_r)\p_+x_{r+1}+\p_+(x_{r-1}-x_r)\p_-
x_{r+1}}{\mbox{ sh }(x_{r-1}-x_r)};\no
\eea
where $k$ is the Cartan matrix of the subalgebra $D_r\in B_r,\; 1\leq i\leq r$;
and possesses the Lagrangian function
\beq
{\it L}=\frac{1}{2}(\sum_{i,j=1}^rk_{ij}\p_+x_i\p_-x_j-
4\mbox{ th }^2\frac{x_{r-1}-x_r}{2}\p_+x_{r+1}\p_-x_{r+1})+
\sum_{i=1}^re^{(kx)_i}.\l{lag}
\eeq
Note that just this nontrivial form of the metric in the target space
causes an origin of a black matter phenomena in the spirit of Witten,
while obtained from nonabelian Toda system, see \cite{GS92}. As it was
also mentioned in that paper by an example of $B_2$, the structure of the
$W$--algebra elements here is highly nontrivial. In particular, one can get
convinced that the form of three most important $W$--elements of the second
order, written before in \cite{S90} only for the case of $B_2$, is
\bea
& &{}^0 W^{\pm}_2=\sum_{i=1}^r\p_{\pm}^2x_i-\frac{1}{2}\sum_{i,j=1}^rk_{ij}
\p_{\pm} x_i\p_{\pm} x_j
+2\mbox{ th }^2\frac{R}{2}(\p_{\pm} x_{r+1})^2;\l{w0}\\
& &{}^{\pm} W^{\pm}_2=e^{\Phi}\p_{\pm} [e^{\pm\Omega -\Phi}(\p_{\pm} R
\pm 2\mbox{ th }\frac{R}{2}
\p_{\pm} x_{r+1})]+e^{\pm\Omega}\p_{\pm} x_{r-2}\p_{\pm} R; \l{w+-}
\eea
where $R\equiv x_{r-1}-x_r,\; \Phi\equiv x_{r-1}+x_r$; and the function
$\Omega$ satisfies the relations
\beq
\p_+\Omega = \frac{\mbox{ ch }R}{\mbox{ ch }^2\frac{R}{2}}\p_+x_{r+1},
\quad
\p_-\Omega = \frac{1}{\mbox{ ch }^2\frac{R}{2}}\p_-x_{r+1}.\l{Omega}
\eeq

Having expressions for the $W$--elements, one can study the problem of the
integrability of a system on a half--line $r\equiv z_+-z_-=0$. Probably, the
simplest way to do that is to equate the corresponding  $W^{\pm}_m$-elements,
$\p_{\mp}W^{\pm}_m=0$, at $z_+=z_-$. For the special case of the Liouville
equation it was shown in \cite{GN82} that its integrability respects the
boundary condition
\beq
\p_r x=\lambda\cdot e^{x} \quad \mbox{at } r=0 \l{bc}
\eeq
for an arbitrary constant $\lambda$. An analogous situation takes place
also for the sine (or sinh)--Gordon model,
see references given in \cite{CDRS94} addressed to a more general
situation with the (abelian) affine Toda field theory associated with an
affine Lie Kac--Moody algebra $\tilde{\cal G}$. Here, except for the
$A_1^{(1)}$, i.e., the sine--Gordon model,  for the simply laced algebras
there are only two
possibilities for constants $\lambda_i$ entering the boundary conditions like
(\ref{bc}) written for the fields $x_i, \; 1\leq i\leq r$; namely
$\vert\lambda_i\vert =2$, or all $\lambda_i =0$. One  can  easily show that
for the corresponding finite case the situation is the same. It
is interesting to note that, in the same way as
for the abelian Toda systems
associated with non simply laced Lie algebras,
 nonabelian Toda systems\cite{BCDR95} in general seemingly
do not obey such  rigid restrictions  on these constants.

Let us show it by the  example of $B_2$ where the boundary conditions contain
one arbitrary constant. Equating the corresponding $W^{\pm}$-elements
(\ref{w0}), (\ref{w+-}), all three of them here being of the $2$nd order,
one has from the expressions for ${}^0W^{\pm}_2$ that
\beq
\p_r x_i=\lambda_i\cdot e^{x_i} \quad \mbox{at } r=0, \; i=1,2; \l{bc1}
\eeq
where $\lambda_i$ are arbitrary constants yet; while ${}^{\pm}W^{\pm}_2$ give
more
rigid conditions that  $\lambda_1= \lambda_2$, and $x_3=\p_r x_3=0$ at $r=0$.
So, finally we have the following, in fact, sufficient boundary conditions:
\beq
\p_r x_i=\lambda\cdot e^{x_i}; \;  i=1,2;\; x_3=\p_r x_3=0 \quad \mbox{at }
r=0.\l{bc2}
\eeq
\section{A B\"acklund Type Transformation}
In this section we discuss a B\"acklund type transformation for the
systems under consideration, and a construction of the corresponding
$W$--elements by an example of equations (\ref{at01.r}) -- (\ref{at-.r4}),
following the line of those in \cite{L84}, \cite{S90}, see also \cite{LS92}.
Let us begin with the connection components of form (\ref{AS});
%\beq
%A_+=g_0^{-1}(\sum_{m=1}^{M-1}V^+_m+X^+)g_0,\quad
%A_-=g_0^{-1}\p_-g_0+\sum_{m=1}^{M-1}V^-_m+X^-;\l{AS}
%\eeq
rewrite them for the case of the series $A_r$ endowed  with
the principal gradation, and take $M=r$. Moreover, supply the
terms $X^{\pm}$ with the factor $\xi$, so to have a possibility
of a direct reduction to the case of the standard Toda
system. Parametrising the functions $g_0$ and $V^{\pm}_m$
as in the section 2, i.e., in our case
\[
g_0=e^{-\sum_{j=1}^rx_jh_j},\quad V^{\pm}_m=
\sum_{p=1}^{r-m+1}e^{-\sum_{j=p}^{m+p-1}
\rho_j}\p_{\pm}f^{\pm m}_p
X_{\pm\sum_{i=p}^{m+p-1}\pi_i};\]
we have
\bea
A_+ &=& \sum_{m=1}^{r-1}\sum_{p=1}^{r-m+1}\p_+f^{+m}_p
X_{+\sum_{i=p}^{m+p-1}\pi_i}+\xi e^{\sum_{j=1}^r\rho_j}
X_{+\sum_{i=1}^r\pi_i},\l{A1}\\
A_- &=& -\sum_{i=1}^r \p_-x_ih_i+
\sum_{m=1}^{r-1}\sum_{p=1}^{r-m+1}e^{-\sum_{j=p}^{m+p-1}
\rho_j}\p_-f^{-m}_pX_{-\sum_{i=p}^{m+p-1}\pi_i}\l{A2}\\
&+&\xi X_{-\sum_{i=1}^r\pi_i}.\no
\eea
Introduce the functions $\Psi^A_a(z_+,z_-)
\equiv <A\vert g\vert a>$, with $\vert a>$ being the basis vectors
of an $N$ dimensional irreducible representation of $A_r$ with the
highest weight components $l_i,\, 1\leq i\leq r$, and $<A\vert$ ---
those of the dual one; and recall that the connection components
$A_{\pm}$ are expressed in terms of $g$ as (\ref{cm}). Since we deal
here with the principal gradation, it is convenient to use the Verma
type basis for the representation space, in other words
\[
\vert a>\equiv \vert j_1,\cdots , j_a>=X_{-j_a}\cdots X_{-j_1}
\vert 0>_l,\; 1\leq j\leq r,\, 0\leq a\leq N-1;\]
$\vert j_1,\cdots , j_0>\equiv \vert 0>_l$;
where $\vert 0>_l$ is the highest weight vector. Thanks to the
defining relation (\ref{dr}),  there take place the following evident
relations:
\bea
& & X_{-j}\vert j_1,\cdots , j_a>_l=\vert j_1,\cdots , j_a, j>_l,\no\\
& & h_j\vert j_1,\cdots , j_a>_l=(l_j-\sum_{b=1}^ak_{j_bj})
\vert j_1,\cdots , j_a>_l,\l{verma}\\
& & X_{+j}\vert j_1,\cdots , j_a>_l=\sum_{b=1}^a\delta_{jj_b}
(l_j-\sum_{c=1}^{b-1}k_{j_cj})\vert j_1,\cdots , j_a>^{(b)}_l;\no
\eea
where the superscript at the vector $\vert j_1,\cdots , j_a>^{(b)}_l$
means that here, in the product $X_{-j_a}\cdots X_{-j_1}$, the
root vector $X_{-j_b}$ is absent. In fact, since we are going to
consider the $1$st fundamental representation of $A_r$, $j_a=a$,
$l_j=\delta_{1j}$, the last formula has the form
\[
X_{+j}\vert a>=\delta_{ja}
(\delta_{1j}-\sum_{c=1}^{a-1}k_{cj})\vert a-1>;\]
here and in what follows the subscript $l$ will be omitted.

Differentiating the functions $<A\vert g\vert a>$ over $z_+$
and over $z_-$ with account of (\ref{cm}), i.e., $\p_{\pm}g=gA_{\pm}$,
and using (\ref{A1}), (\ref{A2}) and (\ref{verma}), after some simple
algebra, we arrive at the equations of the $1$st order for these functions,
\bea
& & \p_+\Psi^A_a=(\p_+f^{+a}_1+\delta_{ar}\xi e^{\sum_{i=1}^r\rho_i})\Psi^A_0
+\sum_{m=1}^{a-1}\p_+f^{+m}_{a-m+1}\Psi^A_{a-m},\l{d1+}\\
& & \p_-(e^{x_1-\sum_{b=1}^a\rho_b}\Psi^A_a)=e^{x_1-\sum_{b=1}^a\rho_b}[
\delta_{a0}\xi \Psi^A_r+\l{d1-}\\
& & \sum_{m=1}^{\mbox{min}(r-a,
r-1)}e^{-\sum_{i=a+1}^{m+a}\rho_i}\p_-f^{-m}_{a+1}\Psi^A_{a+m}].\no
\eea
With the notation $\Phi^A_a\equiv e^{x_1-\sum_{b=1}^a\rho_b}\Psi^A_a$,
the last system (\ref{d1-}) is simplified as
\beq
\p_-\Phi^A_a=\delta_{a0}\xi e^{\sum_{i=1}^r\rho_i}\Phi^A_r+
\sum_{m=1}^{\mbox{min}(r-a, r-1)}\p_-f^{-m}_{a+1}\Phi^A_{a+m}].\l{d2-}
\eeq
Recall that here $0\leq a\leq r$; and $f^0_1=f^r_1\equiv 0$ by definition.

Now one can easily verify that differentiation of system (\ref{d1+}) over $z_-$
and those of system (\ref{d1-}) over $z_+$ gives equations (\ref{at01.r}) --
(\ref{at-.r4}),
so that (\ref{d1+}) and (\ref{d1-}) realise a B\"acklund type transformation
for these equations. From the other hand, with a minor re--notation,
equations (\ref{d1+}) and (\ref{d1-}) are equivalent to the Frenet--Serret
formulas for the moving frame \cite{GM93}; while for other simple Lie algebras
we
expect the similar situation with respect to the $W$--surfaces introduced
in \cite{GS93}. Due to their nilpotent structure, equations (\ref{d1+}) and
(\ref{d1-})
can be explicitly solved in the same way as it is given in \cite{GS94}; their
general solution is expressed in terms of arbitrary functions $\Psi^A_0(z_-)$,
$\Phi^A_r(z_+)$ and the known solution for system (\ref{at01.r}) --
(\ref{at-.r4}). In particular,
the last step in the integration scheme gives the equation of the $(r+1)$th
order for the
functions $\Psi _0^A$, which establishes a relation between its coefficients,
in fact
the $W$--potentials or conserved currents, and the set of linearly independent
fundamental solutions $\Psi ^A_0(z_-)$. Note that such $W$--elements, realised
as differential polynomials in fields $x_i$ and $f^{\pm m}_i$, are nothing but
the characteristic integrals; they  provide the integrability property of
equations (\ref{at01.r})
-- (\ref{at-.r4}) in the same way as the corresponding characteristic integrals
for the standard Toda systems, see e.g., \cite{LSS82}, \cite{L84},
\cite{MSY87}, \cite{LS92}.
To write down the $W$--elements explicitly, let us derive
for them the generating equation, using e.g., system (\ref{d2-}). Since this
equation
is an ordinary differential equation for the function $\Psi_0\equiv
e^{-x_1}\Phi_0$ which
does not depend on the variable $z_+$ thanks to the $1$st equation in
(\ref{d1+}) for $a=0$,
the procedure for obtaining the equation consists in a consequent elimination
of the
functions $\Phi^A_a$ with $a=1$, $a=2$, etc., up to the last one satisfying
$\p_-\Phi^A_r=0$.

One can easily get convinced from (\ref{d2-}) that
\beq
\sum_{a=p}^ru_a\p_-\Phi^A_a=\sum_{a=p+1}^r(\sum_{m=p}^{a-1}u_m
\p_-f^{-(a-m)}_{m+1})\Phi^A_a,\l{su1}
\eeq
with some arbitrary functions $u_a$; and hence
\beq
\sum_{a=p}^r\p_-(u_a\Phi^A_a)=\sum_{a=p+1}^r(\sum_{m=p}^{a-1}u_m
\p_-f^{-(a-m)}_{m+1}+\p_-u_a)\Phi^A_a +\p_-u_p\Phi^A_p.\l{su2}
\eeq
With account of (\ref{su2}) and equations (\ref{d2-}) we obtain the
recurrent relations
\bea
& & \p_-(\frac{1}{v^{(s)}_s}\p_-(\frac{1}{v^{(s-1)}_{s-1}}\p_-(\cdots \p_-(
\frac{1}{v^{(2)}_2}\p_-(\frac{1}{v^{(1)}_1}\p_-\Phi^A_0))\cdots )))\no\\
& & =
v^{(s+1)}_{s+1}\Phi^A_{s+1}+\sum_{a=s+2}^rv^{(s+1)}_a\Phi^A_a;\l{req}
\eea
where
\bea
v^{(n)}_a & = & \sum_{m=n}^{a-1}\frac{v^{(n-1)}_m}{v^{(n-1)}_{n-1}}
\p_-f^{-(a-m)}_{m+1}+\p_-(\frac{v^{(n-1)}_a}{v^{(n-1)}_{n-1}})+
\p_-f^{-(a-n+1)}_n,\quad n\leq a\leq r;\no\\
v^{(1)}_a & = & \p_-f^{-a}_1+\xi \delta_{ar}e^{\sum_{i=1}^r\rho_i},
\quad 1\leq a\leq r;\l{ref}
\eea
and recall once more that $f^{-r}_1\equiv 0$. Finally, one comes to the
following generating equation for the $W$--elements:
\beq
 \p_-(\frac{1}{v^{(r)}_r}\p_-(\frac{1}{v^{(r-1)}_{r-1}}\p_-(\cdots \p_-(
\frac{1}{v^{(2)}_2}\p_-(\frac{1}{v^{(1)}_1}\p_-\Phi^A_0))\cdots )))=0,
\l{geneq}
\eeq
from which the corresponding $W$--elements are constructed in accordance
with the formula
\beq
\sum_{m=0}^rW^-_m\p_-^{r-m+1}\Phi^A_0=0.\l{w-}
\eeq
An analogous representation is easily obtained for the elements $W^+_m$,
$\p_-W^+_m=0$, starting  from system (\ref{d1+}).

Here two comments are in time; the first, equation (\ref{geneq}) should
be understood up to singularity lines related to zeros of the denominators
$v^{(a)}_a$; the second, since not all the fields, contained in $g_0$ and
$V^{\pm}_m$, are the primary ones, the complete set of the $W$--elements
corresponding to system  (\ref{e0.5}) -- (\ref{e-m.5}) or (\ref{at01.r}) --
(\ref{at-.r4}) is, in general, different
than those for the equations satisfied by the primary fields. Of course,
these comments are relevant for nonabelian cases of the standard Toda
systems (\ref{e0.41}) as well. Note also that for the case of the standard
abelian $A_r$ Toda system (\ref{at0.1}), equation (\ref{geneq}) coincides
with those in \cite{L84}.

\section{Outlook}
Finishing up this paper, we would like to formulate  forthcoming
problems which seem to
be
very interesting and important, and can be solved  on the base of the results
obtained
here. First, we have not systematically studied the W-symmetries of
our systems, but we already derived the conserved $W$ charges
 for the
case of $A_r$ in section 8. In our extended  scheme, one may encounter
interesting generalisations of the W-algebras that already arose from
the standard Toda theories. Second,
 introducing matter in the non abelian Toda case is a step towards
 a concrete description of the black hole formation from
collapsing matter and evaporation, using integrable sytem of equations.
Examples of the corresponding field equations were given   in section 6.
Third, it will be interesting to study configurations of the Toda fields
coupled
to  matter fields,
with   boundaries   for abelian and nonabelian
configurations, as was begun in section 7. Fourth,  the considerations  given
in the present paper
can be extended\footnote{This will be done in a
forthcoming article \cite{FGGS95}.}
 in a rather direct way for the case of the systems generated by
the higher grading subspaces of the affine Lie algebras,
 so to have
a generalisation of the abelian and nonabelian affine (conformally affine)
equations and in turn solitons in a black and field matter system; and also
for supersymmetric systems based on the corresponding superalgebras.
Last,  from
the differential geometry point of view it would be very nice to describe,
in the spirit of \cite{GM93}, \cite{GS93}, the $W$--geometries and an analogue
of the Pl\"ucker embedding problem in terms of the solutions to the new systems
discussed in the paper. Moreover, we think that it is possible to realise the
integration
scheme  in terms of some holomorphic and antiholomorphic distributions on the
flag manifolds associated with the relevant parabolic subalgebras of the
algebra
${\cal G}$ in question, in the same way as it has been done for the standard
Toda system (\ref{e0.41}) in \cite{RS94}.

\section{Acknowledgements}
One of the authors (M. S.) wishes to acknowledge the warm
hospitality and creative scientific atmosphere of the
Laboratoire de Physique
Th\'eorique de l'\'Ecole
Normale Sup\'erieure de Paris. This work was partially supported by the
Russian Fund for Fundamental Research and International Science Foundation.

\end{document}